\documentclass[a4paper,12pt]{article}
\usepackage{amsmath}
\usepackage{amsfonts}
\usepackage{amssymb}
\usepackage{graphicx} 

\title{\textsc{\begin{LARGE}Gravity and Large Extra Dimensions\end{LARGE}}}
\author{\textbf{$\hbox{V H Satheesh Kumar}^{\star}$} and \textbf{{$\hbox{P K Suresh}^{\dagger}$}} \\ School of Physics, University of Hyderabad \\ Hyderabad 500 046, India. \\ $^\star$ {\begin{small}\texttt{vhsatheeshkumar@yahoo.co.uk}\end{small}}, $^\dagger$ \begin{small}\texttt{pkssp@uohyd.ernet.in}\end{small}}
\date{}

\begin{document}
\maketitle
\begin{abstract}

The idea that quantum gravity can be realized  at the TeV scale is extremely attractive to theorists and experimentalists alike. This proposal leads to extra spacial dimensions large compared to the electroweak scale. Here we give a very systematic view of the foundations of the theories with large extra dimensions and their physical consequences.

\end{abstract}

\newpage

\section{Introduction}

The idea of extra dimensions slipped into the realm of physics in the 1920's when Kaluza and Klein \cite{KKT} tried to unify electromagnetism with gravity, by assuming that the electromagnetic field originates from fifth component ($g_{\mu 5}$) of a five dimensional metric tensor. The development of string theory in early 1980's lead to a revitalization of the idea of extra dimensions. 

The first indication of large extra dimensions in string theory came in 1988 from studies of the problem of supersymmetry breaking by Antoniadis \textit{et Al} \cite{Antoniadis:1988jn}. Supersymmetry was introduced to make the masses of elementary particles compatible with graviton. Quantum gravity without supersymmetry introduces a new scale, the Planck mass $\sim 10^{19}$ Ge, which is $10^{16}$ times heavier than the observed electroweak scale. This is the so-called mass hierarchy problem. Since no superparticle, as predicted by supersymmetry, has ever been produced in accelerator, they must be heavier than the observed particles. Supersymmetry therefore should be broken. On the other hand, protection of mass hierarchy requires that its breaking scale cannot be larger than a few TeV. Assuming that supersymmetry breaking in string theory arises by the process of compactification of the extra dimensions, Antoniadis  \cite{antoniadis} showed that its energy breaking scale is tied to the size $10^{-18}m$. There was little interest in such models with large dimensions because of theoretical reasons related to the large string coupling problem.

In 1996, Witten \cite{Witten:1996mz} proposed that the string size is a free parameter of the theory, with \textit{a priori} no relation to the Planck length. In particular, it could be as large as $10^{-18}$ m  which is just below the limiting distance that can be probed by present experiments \cite{Lykken:1996fj}. Since, with the advent of many duality symmetries, computations with large coupling became effectively possible, the road was open to study models with extra dimensions much larger than the Planck length.

In 1998, Arkani-Hamed, Dimopoulos and Dvali (ADD) \cite{ADD} turned down the approach to the hierarchy problem by introduction of supersymmetry at electroweak energies (compactification at the electroweak scale were first considered in  \cite{Volobuev:1986wc}).  Rather than worrying about the inconvenient size of the Plank length, they wondered what gravity would look like if it too operated at electroweak scale, making it stronger than we realize. 
The problem is solved by altering the fundamental Planck scale with the help of $n$ new spatial dimensions large compared to the electroweak scale. The most attractive feature of this framework is that it is not experimentally excluded like string theory. Firstly, quantum gravity has been brought down from $10^{19}$ Ge to $\sim$ TeV. Secondly,   the structure of spacetime has been drastically modified at sub-mm distances. Thus, it gives rise to new predictions that can be tested in accelerator, astrophysical and table-top experiments \cite{ADDphen}.  Moreover, the framework can be embedded in string theory \cite {AADD}. However, currently the only non-supersymmetric string models that can realize the extra dimensions and break to only the standard model particles at low energy with no extra massless matter are \cite{Cremades:2002dh, Kokorelis:2002qi}.

The simplest ADD scenario is characterized by the SM fields localized on a four dimensional submanifold of thickness $m_{EW}^{-1}$ in the extra $n$ dimensions, while gravity spreads to all $4+n$ dimensions. The $n$ extra dimensions are compactified and has a topology  $R^4 \times M_n$,  where $M_n$ is an $n$ dimensional compact manifold of volume $R^n$. All extra dimensions have equal size $L=2 \pi R$. The fundamental scale of gravity $M_*$, and the ultraviolet scale of the Standard Model, are around a few TeV or so. The $(4+n)$ dimensional Planck mass is $\sim m_{EW}$, the only short-distance scale in the theory. Therefore the gravitational force becomes comparable to the gauge forces at the weak scale.  

In this short review we would like present most of the features of ADD model. Other recent reviews of the subject can be found in \cite{LEDreviews}. The interested readers can refer to \cite{Duff} for good reviews on Kaluza-Klein Theories and a very good introduction to extra dimensions can be found in \cite{VHS}. 

\section{Localization}

Why are any of the standard model particles or fields, in any experiment so far conducted are not disappearing into extra dimensions? Answering this question will naturally lead us towards theories with SM matter and fields localized to branes. Then the new question arises, what is the mechanism by which the standard model fields are localized to the brane? The idea of localizing particles on walls (brane) in a higher dimensional space goes back to Akama \cite{Akama}, Rubakov and Shaposhnikov \cite{RubShap} and  Visser \cite{Visser} whose ideas relied on the index theorem in soliton background \cite{RS}. 

\subsection{Fermions}

Rubakov and Shaposhnikov \cite{RubShap} constructed the first field theoretic models with localized fermions. ADD \cite{ADD} generalized this to include their framework. Massless four-dimensional fermions localized on the domain wall (zero modes) are meant to mimic SM particles. They acquire small masses through the usual Higgs mechanism.  Explicit expressions for fermion zero modes in various backgrounds are given in \cite{Jackiw:1981ee,'tHooft:1976fv, DvaliShifman}. At low energies, their interactions can produce only zero modes again, so physics is effectively four-dimensional. This possibility of explaining the origin of three Standard Model generations has been explored in \cite{Libanov:2001uf,Frere:2000dc}. Zero modes interacting at  high energies, however, will produce continuum modes,  the extra dimension will open up, and particles will be able to leave the brane, escape into  extra dimension and literally disappear from our world. For a four-dimensional observer these high energy processes will look like $e^+ e^- \to$  \textit{nothing} or $e^+ e^- \to  \gamma +$  \textit{nothing}. We shall discuss later how these and similar processes are indeed possible in accelerators and are used to probe the existence of extra dimensions.

\subsection{Gauge Fields}

A mechanism for gauge field localization within the field theory context  was proposed by Dvali and Shifman \cite {DS}. It is based on the observation that gauge field can be in the  confining phase on the bulk while being in the broken phase on a brane;  then confining potential prevents the low energy  brane gauge fields to propagate into the bulk. This has been generalized to higher dimensions in \cite{ADD}. Antoniadis, Arkani-Hamed, Dimopoulos and Dvali \cite{AADD} have shown that this framework can naturally be embedded in type I string theory. This has the obvious advantage of being formulated within a consistent theory of gravity, with the additional benefit that the localization of gauge theories on a three brane is automatic  \cite{Dbrane}. Further interesting progress towards realistic string model building was made in \cite{Tye}.

The standard model fields are only localized on the brane of width  $M_{*}^{-1}$ in the bulk of $4+n$ dimensions. In sufficiently hard collisions of energy $E_{esc} \gtrsim M_{*}$, they can acquire momentum in the extra dimensions and escape from our four dimensional world, carrying away energy. Usually  in theories with extra compact dimensions of size $R$, states with momentum in the compact dimensions are interpreted from the four dimensional point of view as particles of mass $1/R$, but still localized in the four dimensional world. This is because, at the energies required to excite these particles, there wavelength and the size of the compact dimension are comparable. In ADD case the situation is completely different: the particles which can acquire momentum in the extra dimensions have TeV energies, and therefore have wavelengths much smaller than the size of the extra dimensions. Thus, they simply escape into the extra dimensions. In fact, for energies above the threshold $E_{esc}$, escape into the extra dimensions is enormously favored by phase space. This implies a sharp upper limit to the transverse momentum which can be seen in 4 dimensions at $p_T = E_{esc}$, which may be seen at accelerators if the beam energies are high enough to yield collisions with center-of-mass energies greater than $E_{esc}$.  

Notice that while energy can be lost into the extra dimensions, electric charge (or any other unbroken gauge quantum number) can not be lost. This is because  the massless photon is localized in our Universe and an isolated charge can not exist in the region where electric field can not penetrate, so charges can not freely escape into the bulk, although energy may be lost in the form of neutral particles propagating  in the bulk. Similar conclusions can be  reached by considering a soft photon emission process in \cite{Barnaveli:1990pm}. Once the particles escape into the extra dimensions, they may or may not return to the four dimensional world, depending on the  topology of the $n$ dimensional compact manifold $M_n$.  In the most interesting case, the particles orbit around the extra dimensions,  periodically returning, colliding with and depositing  energy to our four dimensional world with  frequency $R^{-1}$.

\section{Relating Plank Scales}

The important question that we would like to answer is how large the extra dimensions could possibly be without us having them noticed until now. For this we need to understand how the effectively four dimensional world that we observe would be arising from the higher dimensional theory. Let us call the fundamental (higher dimensional) Planck scale of the theory $M_*$, assume that there are $n$ extra dimensions, and that the radii of the extra dimensions are given by $R$. We will carry out this simple exercise in three different ways. 

\subsection{Gauss law}

The easiest derivation is a trivial application of Gauss' Law. The $(4+n)$ dimensional Gauss' law for gravitational interaction is given by
\begin{equation}
\left(\begin{array}{c}
\mbox{Net gravitational flux}\\
\mbox{over a closed surface C}
\end{array}\right)\ = S_{d}\, G_{N(4+n)} \times
\mbox{Mass in}\, C
\end{equation}
where $S_{d} = 2 \pi^{d/2} / \Gamma(d/2)$ is the surface area of the unit sphere in $d$ spatial dimensions. 

Suppose now that a point mass $m$ is placed at the origin. One can reproduce this situation in the uncompactified theory by placing ``mirror" masses periodically in all the new dimensions. Of course for a test mass at distances $r \ll L$ from $m$, the ``mirror" masses make a negligible small contribution to the force and we have the $(4+n)$ dimensional force law,
\begin{equation}
F_{(4+n)}(r) = G_{N(4+n)} \frac{m_1 m_2}{r^{n+2}}.
\end{equation} 

For $r \gg L$, on the other hand, the discrete distance between mirror masses can not be discerned and they look like an infinite $n$ spatial dimensional ``line" with uniform mass density. The problem is analogous to  finding the gravitational field of an infinite line of mass with uniform mass/unit length, where cylindrical symmetry and Gauss' law give the  answer. Following exactly the same procedure, we consider a ``cylinder" $C$ centered around the $n$ dimensional line of mass, with side length $l$ and end caps being three dimensional sphere's of radius $r$. In our case, the LHS is equal to $F(r) \times 4 \pi \times l^n$, while the total mass contained in $C$ is $m \times (l^n/L^n)$. Equating the two sides, we find the correct $1/r^2$ force law and can identify
\begin{equation}
G_{N(4)} = \frac{S_{(3+n)}}{4 \pi} \frac{G_{N(4+n)}}{V_n}
\label{GN}
\end{equation}
where $V_n = L^n$ is the volume of compactified dimensions.

 The two test masses of mass $m_1,m_2$ placed within a distance $r \ll L$ will feel a gravitational potential dictated by Gauss's law in $(4+n)$ dimensions
\begin{equation}
V(r) \sim \frac{m_1 m_2}{M_{*}^{n+2}} \frac{1}{r^{n+1}}, \, (r \ll L).
\end{equation}
On the other hand, if the masses are placed at distances $r \gg L$, their
gravitational flux lines can
not continue to penetrate in the extra dimensions, and the usual $1/r$
potential is obtained,
\begin{equation}
V(r) \sim  \frac{m_1 m_2}{M_{*}^{n+2} R^{n}} \frac{1}{r}, \, (r \gg L)
\end{equation}
so our effective 4 dimensional $M_{Pl}$ is
\begin{equation}
M_{Pl}^2 \sim M_{*}^{2 + n} R^n. \label{fourplanck}
\end{equation}

\subsection{Action Method}

In this method, we first write down the action for the higher dimensional gravitational theory, including the dimensionful constants and then dimensionally reduse it to compare the quantities. Here, we use the mass dimensions of the various quantities for analysis.

The higher dimensional line element is given by 
\begin{equation}
ds^2= g_{\hat{\mu} \hat{\nu}} dx^{\hat{\mu}} dx^{\hat{\nu}}
\end{equation}
The corresponding Einstein-Hilbert action in $n$ dimensions can be written as 
\begin{equation}
S_{(4+n)} \sim \int d^{(4+n)}x \sqrt{|g^{(4+n)}|} {\cal{R}}^{(4+n)}.
\end{equation}
In order to make the action dimensionless, we need to multiply by the appropriate power of the fundamental Planck scale $M_*$. Since $d^{(4+n)} x$ carries dimension $-n-4$ and ${\cal{R}}^{(4+n)}$ carries dimension $2$, $M_*$ should have the power $n+2$, thus 
\begin{equation}
\label{extradaction}
S_{(4+n)} = -M_*^{(n+2)} \int d^{(4+n)}x \sqrt{|g^{(4+n)}|} {\cal{R}}^{(4+n)}.
\end{equation}
While the usual four dimensional action is given by
\begin{equation}
\label{4daction}
S_{(4)} = -M_{Pl}^{2} \int d^{(4)}x \sqrt{|g^{(4)}|} {\cal{R}}^{(4)},
\end{equation}
where $M_{Pl}$ is the observed four dimensional Planck scale $\sim 10^{18}$ GeV.

Now, to compare these two actions, we assume that spacetime is flat and that the $n$ extra dimensions are compact. So the $n$ dimensional metric is given by
\begin{equation}
\label{metric}
ds^2 = (\eta_{\mu\nu}+h_{\mu\nu}) dx^\mu dx^\nu -R^2 d \Omega_{(n)}^2,
\end{equation}
where $x_\mu$ is a four dimensional coordinate, $ d \Omega_{(n)}^2$ corresponds to the line element of the flat extra dimensional space in some parameterization, $\eta_{\mu\nu}$ is the four dimensional Minkowski metric, and $h_{\mu\nu}$ is the four dimensional fluctuation of the metric around its minimum. From this we calculate the expressions 
\begin{equation}
\sqrt{|g^{(4+n)}|} = R^n \sqrt{|g^{(4)}|}, \ \ {\cal{R}}^{(4+n)}= {\cal{R}}^{(4)}.
\end{equation}
Substituting these quantities in (\ref{extradaction}), we get 
\begin{equation}
S_{(4+n)} = -M_*^{n+2} \int d \Omega_{(n)} R^n \int d^{(4)}x \sqrt{|g^{(4)}|} {\cal{R}}^{(4)}.
\end{equation} 
The factor $\int d \Omega_{(n)} R^n$ is nothing but the volume of the extra dimensional space which we denote by $V_{(n)}$. For toroidal compactification it would simply be  given by $V_{(n)} =(2\pi R)^n$. Therefore the above action takes the form,
\begin{equation}
\label{matchingaction}
S_{(4+n)} = -M_*^{n+2}  (2\pi R)^n \int d^{(4)}x \sqrt{|g^{(4)}|} {\cal{R}}^{(4)}.
\end{equation} 
Comparing (\ref{matchingaction}) with (\ref{4daction}) we find
\begin{equation}
M_{Pl}^2= M_*^{n+2} (2\pi R)^n.
\label{grmatching}
\end{equation}

\subsection{Kaluza-Klein Method}

Finally, we can understand this result purely from the 4-dimensional point of view as arising from the sum over
the Kaluza-Klein excitations of the graviton. From the 4-d point of view, a $(4+n)$ dimensional graviton with momentum $(p_1, \cdots, p_n)$ in the extra $n$ dimensions looks like a massive particle of mass $|p|$. Since the
momenta in the extra dimensions are quantized in units of $1/R$, this corresponds to an infinite tower of KK excitations for each of the $n$ dimensions, with mass splittings $1/R$. While each of these KK modes is very
weakly coupled ($\sim 1/M_{(4)}$), their large multiplicity can give a large enhancement to any effect they mediate. In our case, the potential between two test masses not only has the $1/r$ contribution from the usual massless
graviton, but also has Yukawa potentials mediated by all the massive modes as well:
\begin{equation}
\frac{V(r)}{m_1 m_2} = G_{N(4)} \sum_{\vec{n}}
\frac{e^{{-(|\vec{n}|}/{R}) r}}{r}.
\end{equation}
Obviously, for $r \ll L$, only the ordinary massless graviton  contributes and we have the usual potential. For $r \gg L$, however, roughly $(L/r)^n$ KK modes make unsuppressed contributions, and so the potential grows more rapidly
as $L^n/r^{n+1}$. More exactly, for $r \gg L$,
\begin{eqnarray}
\frac{V(r)}{m_1 m_2} &\to&  {G_{N(4)}} r \times {\left( \frac{L}{2 \pi r}\right) }^n
\times
\int d^n u e^{-|u|} \nonumber \\
&=& \frac{G_{N(4)} V_n}{r^{n+1}} \times \frac{S_n \Gamma(n)}{(2 \pi)^n}.
\end{eqnarray}
Upon using the Legendre duplication formula
\begin{equation}
\Gamma{\left( \frac{n}{2}\right) } \Gamma{\left( \frac{n}{2}\right) } + \frac{1}{2} =
\frac{\sqrt{\pi}}{2^{n-1}} \Gamma(n),
\end{equation}
this yields the same relationship between $G_{N(4)}$ and $G_{N(4+n)}$ as found earlier
\begin{equation}
M_{Pl}^2= M_*^{n+2} (2\pi R)^n.
\end{equation}

\section{Size of Extra Dimensions}

An important issue in extra dimensional theories is the mechanism by which  extra dimensions are hidden, so that the spacetime is effectively four dimensional insofar as known physics is concerned. The most plausible way of achieving this is by assuming that these extra dimensions are finite and are compactified. Then one would need to be able to probe length scales corresponding to the size of the extra dimensions to be able to detect them. If the size of the extra dimensions is small, then one would need extremely large energies to be able to see the consequences of the extra dimensions. Thus by making the size of the extra dimensions very small, one can effectively hide these dimensions. So the most important question that one needs to ask is how large could the size of the extra dimensions be without getting into conflict with observations?

The new physics will only appear in the gravitational sector  when distances as short as the size of the extra dimension are actually reached.  However, it is very hard to test gravity at very short distances. The reason is that gravity is a much weaker interaction than all the other forces. Over large distances gravity is dominant, however, as one starts going to shorter distances, inter-molecular van der Waals forces and eventually bare electromagnetic forces will be dominant, which will completely overwhelm the gravitational forces. This is the reason why  the Newton-law of gravitational interactions  has only been tested down to about a fraction of a millimeter using  essentially Cavendish-type experiments~\cite{gravityexp}. Therefore, the real bound on the size of an extra dimension is $R \leq 0.1\ {\rm mm}$, if only gravity propagates in the extra dimensions. How would a large value close to the experimental bound affect the fundamental Planck scale $M_*$? Since we have the relation $M_{Pl}^2 \sim M_*^{n+2} R^n$, if $R> 1/M_{Pl}$, the fundamental Planck scale $M_*$ will be lowered from $M_{Pl}$. How low could it possibly go down? If $M_* < 1$ TeV, that would imply that quantum gravity should have already played a role in the collider experiments that  have been performed until now. Since we have not seen a hint of that,  one has to impose that $M_* \geq 1$ TeV. So the lowest possible  value, and thus the largest possible size of the extra dimensions, would be  for $M_* \sim 1$ TeV.

Let us check, how large a radius one would need, if in fact $M_*$ was of the order of a TeV. Reversing the expression  $M_{Pl}^2 \sim M_*^{n+2} R^n$ we would now get 
\begin{equation}
\frac{1}{R}=M_* \left( \frac{M_*}{M_{Pl}}\right)^\frac{2}{n} = 
(1 {\rm TeV}) 10^{-\frac{32}{n}},
\end{equation}
where we have used $M_* \sim 10^3$ GeV and $M_{Pl} \sim 10^{19}$ GeV. To convert into conventional length scales one should keep the  conversion factor $1 {\rm GeV}^{-1} =2 \times 10^{-14} {\rm cm}$ in mind. Using this we finally get
\begin{equation}
R\sim 2\times 10^{-17} \times 10^{\frac{32}{n}} \ {\rm cm}.
\end{equation}

For $n=1$, $R \sim 10^{13}$ cm, this case is obviously excluded since it would modify Newtonian gravitation at solar-system distances. However, already for two extra dimensions one would get a much smaller number $R\sim 2$ mm. This is  just borderline excluded by the latest gravitational experiments performed in Seattle~\cite{Eotwash}. Conversely, one can set a bound on the size of two large extra  dimensions from the Seattle experiments, which gave $R\leq 0.2$ mm$=10^{12}$ 1/GeV. This results in $M_* \geq 3$ TeV. We will see that for two extra  dimensions there are in fact more stringent bounds than the  direct bound from gravitational measurements.

For $n>2$ the size of the extra dimensions is less than $10^{-6}$ cm, which is unlikely to be tested directly via gravitational measurements any time soon. Thus for $n>2$ $M_* \sim 1$ TeV is indeed a possibility that one has to 
carefully investigate. If $M_*$ was really of order the TeV scale, there  would no longer be a large hierarchy between the  fundamental Planck scale $M_*$ and the scale of weak interactions $M_{EW}$, thus this would resolve the 
hierarchy problem. In this case gravity would appear weaker than the other forces at long distances because it would get diluted by the large volume of the extra dimensions. However, this would only be an apparent hierarchy  between the strength of the forces, as soon as one got below scales of order $r$ one would start seeing the fundamental gravitational force, and the  hierarchy would disappear. However, as soon as one postulates the equality  of the strength of the weak and gravitational interactions one needs to  ask why this is not the scale that sets the size of the extra dimensions  themselves. Thus by postulating a very large radius for the extra dimensions one would merely translate the hierarchy problem of the scales of interactions  into the problem of why the size of the extra dimension is so large compared  to its natural value.

\section{Graviton Spectrum}

In this section, we compactify D dimensional gravity on an n-dimensional torus and perform mode expansion. The graviton corresponds to the excitations of the $D$-dimensional metric. In terms of 4-dimensional indices, the metric tensor contains spin-2, spin-1, and spin-0 particles.  Moreover, since these fields depend on $D$-dimensional coordinates, they  can be expressed as a tower of Kaluza-Klein modes. The mass of each Kaluza-Klein mode corresponds to the modulus of its momentum in the direction transverse to the brane. The picture of a massless graviton propagating
in $D$ dimensions and the picture of massive Kaluza-Klein gravitons propagating in 4 dimensions are equivalent.

The 4+n dimensional metric is given by
\begin{equation}
\hat{g}_{\hat{\mu} \hat{\nu}}=\eta_{\hat{\mu}\hat{\nu}}+\hat{k} \hat{h}_{\hat{\mu}\hat{\nu}},
\end{equation} 
where $\hat\kappa^2=16\pi G_N^{(4+n)}$, with $G_N^{(4+n)}$ the Newton constant in $D=4+n$. The Einstein-Hilbert action for the above metric can be written as,
\begin{equation}
{\hat S}= \frac{1}{\hat{k}^2}\int d^{4+n}\hat{x} \sqrt{|{\hat g}|} {\hat{\cal{R}}}
\end{equation}
where $\hat{\cal{R}}$ is the 4+n dimensional curvature invariant. This action is invariant under the 4+n dimensional general coordinate transformations,
\begin{equation}
\delta \hat{h}_{\hat{\mu} \hat{\nu}} = \partial_{\hat{\mu}} \zeta_{\hat{\nu}} + \partial_{\hat{\nu}} \zeta_{\hat{\mu}}.
\end{equation}

Clearly, the graviton is a $D\times D$ symmetric tensor, where $D=4+n$ is the total number of dimensions. Therefore this tensor has in principle $D(D+1)/2$ components. However, because $D$ dimensional general coordinate invariance, we can impose $D$ separate conditions to fix the gauge, for example, using the harmonic gauge 
\begin{equation}
\partial_{\hat{\mu}} h^{\hat{\mu}}_{\hat{\nu}} =\frac{1}{2} \partial_{\hat{\nu}} h^{\hat{\mu}}_{\hat{\mu}}.
\end{equation}
This brings down the number of degrees of freedom by D. However, this is not yet a complete gauge fixing. Gauge transformations which satisfy the equation  $\Box \epsilon_{\hat{\mu}}=0$ are still allowed, where the gauge transformation is 
\begin{equation}
h_{{\hat{\mu}}{\hat{\nu}}}\to h_{{\hat{\mu}}{\hat{\nu}}}+\partial_{\hat{\mu}} \epsilon_{\hat{\nu}}+ \partial_{\hat{\nu}} \epsilon_{\hat{\mu}}, 
\end{equation} 
and this means that another $D$ conditions can be imposed. This means that generically a graviton has  $D(D+1)/2-2D=D(D-3)/2$ independent degrees of freedom. For $D=4$ this gives the usual 2 helicity states for a massless spin-two particle, however in $D=5$ we get 5 components, in $D=6$ we get 9 components, etc. This means that from the four dimensional point of view a higher dimensional graviton will contain particles other than just the ordinary four dimensional graviton. 

Now let us discuss the different modes D dimensional graviton from four dimensional perspective. To perform the KK reduction,  we shall assume 
\begin{equation}
\hat{h}_{\hat{\mu} \hat\nu}\ =\ 
V_n^{-1/2}\left(\begin{array}{cc}
h_{\mu\nu}+\eta_{\mu\nu}\phi & A_{\mu j}\\
A_{i \nu}   &  2 \phi_{ij}
\end{array}\right)\ ,
\end{equation}
where $V_n$ is the volume of the $n$ extra dimensional compactified space, $\phi\equiv\phi_{ii}$, $\mu, \nu=0,1,2,3$ and $i, j=5,6,\cdots ,4+n$. These fields are compactified on an $n$-dimensional torus $T^n$ and have the following mode expansions,
\begin{eqnarray}
h_{\mu\nu}(x,y)\ &=&\ \sum_{\vec n} h_{\mu\nu}^{\vec{n}}(x)\ 
\exp\left(i {2\pi \vec{n}\cdot\vec{y}\over R}\right)\ ,\\
A_{\mu i}(x,y)\ &=&\ \sum_{\vec n} A_{\mu i}^{\vec{n}}(x)\ 
\exp\left(i {2\pi \vec{n}\cdot\vec{y}\over R}\right)\ ,\\ 
\phi_{ij}(x,y)\ &=&\ \sum_{\vec n} \phi_{ij}^{\vec{n}}(x)\ 
\exp\left(i {2\pi \vec{n}\cdot\vec{y}\over R}\right)\ , \\
{\vec n} &=& \{n_1,n_2,\cdots,n_n\}\ ,\label{mode}
\end{eqnarray}
where the modes of $\vec{n}\neq 0$ are the KK states, and all the compactification radii are assumed  to be the same.  From the transformation properties under the general coordinate transformation $\zeta_{\hat{\mu}}=\{\zeta_\mu, \zeta_i\}$, it should be clear that the zero modes, $\vec{n}=\vec{0}$,  correspond to the massless graviton,  gauge bosons and scalars in four dimensions.

The above KK modes satisfy the following equation of motions
\begin{eqnarray}
(\Box + m^2_{\vec{n}})\ (h^{\vec{n}}_{\mu\nu}-{1\over2}\eta_{\mu\nu} h^{\vec{n}})\ &=&\ 0,\\
(\Box + m^2_{\vec{n}})\ A^{\vec{n}}_{\mu i}\ &=&\ 0,\\
(\Box + m^2_{\vec{n}})\ \phi_{ij}^{\vec{n}}\ &=&\ 0,
\end{eqnarray}
where $\Box$ is the four-dimensional d'Alembert operator and $m$ is the mass of the graviton mode given by,
\begin{equation}
m^2_{\vec{n}}\ =\ {4\pi^2 {\vec{n}}^2\over R^2}.
\end{equation} 

The different four dimensional fields are coming from the different blocks in the bulk graviton metric, which is represented aesthetically below.
\begin{equation}
\left( \begin{array}{c|ccc} 
h^{\vec{n}}_{\mu\nu} & & A^{\vec{n}}_{\mu j}\\ 
\hline 
& & &  \\
A^{\vec{n}}_{i \nu}& & \phi_{ij}^{\vec{n}} \\ & & &  \end{array} \right)
\end{equation}

\subsection{Zero Modes}

The bulk graviton is given by a $(4+n)\times(4+n)$ matrix. The zero mode of four dimensional graviton comes from the upper left $4 \times 4$ block. There is only one such massless spin-2 particle (graviton) with  two degrees of freedom. The off-diagonal blocks of the bulk graviton form vectors under the four dimensional Lorentz group. Since there are $n$ such vectors, we have $n$ corresponding massless spin-1 particles (vector gauge bosons), each with two degrees of freedom. The remaining lower right $n \times n$ block of the bulk graviton matrix clearly corresponds to four dimensional scalar fields. This has $n(n+1)/2$ spin-0 particles (scalars) corresponding to each independent term of the $n \times n$ matrix. Each scalar has one degree of freedom. Summing all the degrees of freedom we get $2+2n+n(n+1)/2$, which is precisely the total number of degrees of freedom that $4+n$ dimensional graviton can have.

\subsection{Kaluza-Klein Modes}

For non-zero modes the upper left $4 \times 4$ block represents a massive spin-2 particle (massive graviton) with five degrees of freedom. The reason is that a massive graviton contains a normal four dimensional massless graviton with two components, but also ``eats'' a massless gauge  field and a massless scalar, as in the usual Higgs mechanism. Thus $5=2+2+1$. Earlier we had $n$ massless gauge bosons and we are left with only $n-1$ as one of it is eaten away by the graviton. Now each of these vectors absorb a scalar via the Higgs mechanism and become massive and have three degrees of freedom each. Now there are only $n(n-1)/2$ massive scalars, each with one degree of freedom. Summing all the degrees of freedom of these non-zero modes, we get $5+3(n-1)+n(n-1)/2$, which is precisely the total number of degrees of freedom that $4+n$ dimensional graviton can have.

\section{Coupling of the KK States to SM Fields}

In this section we would like to explicitly construct the generic interaction Lagrangians between the matter on the brane and the various graviton modes. For our discussion we will follow the work of Giudice, Rattazzi and Wells \cite{GRW} and Han, Lykken and Zhang \cite{HLZ}.

The action with minimal gravitational coupling of the general scalar $\Phi$, vector $A$, and fermion $\Psi$ is given by 
\begin{equation}
S=\int d^4x  \sqrt{-\hat{g}} {\cal L}_{SM}(\hat{g}_{{\mu} {\nu}}, \Phi ,\Psi ,A)
\label{coup}
\end{equation}
The ${\cal O}(\kappa)$ term of Eq.~(\ref{coup}) can be easily shown to be
\begin{equation}
S=-{\kappa\over 2} \int d^4 x (h^{\mu\nu} T_{\mu\nu}+ \phi T^\mu_{~\mu})\ ,
\end{equation}
where 
\begin{equation}
T_{\mu\nu}(\Phi ,\Psi ,A)\ =\ \biggl(-{\eta_{\mu\nu}}{\cal L}+2 {\delta{\cal L}\over\delta\hat{g}^{\mu\nu}} 
\biggr)\vert_{\hat{g}=\eta}\ , 
\end{equation}
and we have used 
\begin{eqnarray}
\sqrt{-\hat{g}} &=& 1+{\kappa\over2}h + 2\kappa\phi\ , \\ 
\hat{g}_{\mu\nu} &=&\eta_{\mu\nu} + \kappa (h_{\mu\nu} + \eta_{\mu\nu} \phi), \\
\hat{g}^{\mu\nu}&=&\eta^{\mu\nu}-\kappa h^{\mu\nu}-\kappa\eta^{\mu\nu}\phi\ ,\\
\phi&\equiv&\phi_{ii}, \\
\kappa&=&\sqrt{16\pi G_N}, \\
\kappa&=&V_n^{-1/2}\hat\kappa,\\
V_n&=&R^n.
\end{eqnarray} 

For the KK modes, we replace ${h^{\vec{n}}_{\mu\nu}} \hbox{and} \phi^{\vec{n}}$  by the physical fields ${\tilde{h}}^{\vec{n}}_{\mu\nu} \hbox{and} {\tilde{\phi}}^{\vec{n}}$. This redefinition of the fields is associated with spontaneous symmetry breaking, whose details are not given here. With these new quantities action takes the form,
\begin{equation}
S=-{\kappa\over 2} \sum_{\vec{n}} \int d^4 x (\tilde{h}^{\mu\nu,\vec{n}} T_{\mu\nu} + \omega\tilde{\phi}^{\vec{n}} T^\mu_{~\mu})\ .\label{gen}
\end{equation}
where $\tilde{\phi}^{\vec{n}}\ \equiv\ \tilde{\phi}^{\vec{n}}_{ii}$ and $\omega=\sqrt{2/3(n+2)}$.

In the following, we present only three-point vertex Feynman rules and the energy-momentum tensor for scalar bosons, gauge bosons and fermions where we have used the following symbols,
 \begin{eqnarray}
 &&C_{\mu\nu,\rho\sigma}\ =\ \eta_{\mu\rho}\eta_{\nu\sigma}
+\eta_{\mu\sigma}\eta_{\nu\rho}
-\eta_{\mu\nu}\eta_{\rho\sigma}\ ,
\label{C}\\
&&
D_{\mu\nu,\rho\sigma} (k_1, k_2)\ =\
\eta_{\mu\nu} k_{1\sigma}k_{2\rho}
- \biggl[\eta_{\mu\sigma} k_{1\nu} k_{2\rho}
  + \eta_{\mu\rho} k_{1\sigma} k_{2\nu}
  - \eta_{\rho\sigma} k_{1\mu} k_{2\nu}
  + (\mu\leftrightarrow\nu)\biggr]\ ,\nonumber\\
&&\label{D}\\
&&
E_{\mu\nu,\rho\sigma} (k_1, k_2)\ =\ \eta_{\mu\nu}(k_{1\rho}k_{1\sigma}
+k_{2\rho}k_{2\sigma}+k_{1\rho}k_{2\sigma})
\nonumber\\
&&\qquad\qquad\qquad\qquad
-\biggl[\eta_{\nu\sigma}k_{1\mu}k_{1\rho}
+\eta_{\nu\rho}k_{2\mu}k_{2\sigma}
+(\mu\leftrightarrow\nu)\biggr].
\end{eqnarray}

The four-point and five-point vertex Feynman rules and their derivation can be found in \cite{HLZ}.

\subsection{Scalar Bosons}
The conserved energy-momentum tensor for scalar bosons is
\begin{equation}
T^{\rm S}_{\mu\nu}\ =\ -\eta_{\mu\nu}D^\rho\Phi^\dagger D_\rho\Phi+\eta_{\mu\nu}m^2_\Phi\Phi^\dagger\Phi +D_\mu\Phi^\dagger D_\nu\Phi+D_\nu\Phi^\dagger D_\mu\Phi\ ,
\end{equation}
where the gauge covariant derivative is defined as $D_\mu = \partial_\mu + i g A^a_\mu T^a,$ with $g$ the gauge coupling, $A^a_\mu$ the gauge fields and $T^a$ the Lie algebragenerators. 
\begin{center}
\includegraphics[scale=1]{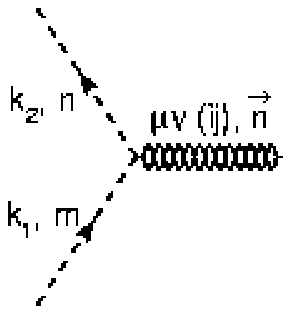}
\end{center}

\begin{eqnarray}
{\tilde{h}}_{\mu \nu}^{\vec{n}}\Phi\Phi&:&\, \, \,-i\, \kappa/2 \, \delta_{mn}(m^2_\phi \eta_{\mu \nu}+C_{\mu\nu,\rho\sigma}k_1^\rho k_2^\sigma)\\
{\tilde{\phi}}^{\vec{n}}_{ij}\Phi\Phi&:&\, \, \,i\,\omega \, \kappa \,\delta_{ij} \delta_{mn}( k_1 \cdot k_2-2 m^2_\phi)
\end{eqnarray}

\subsection{Gauge Bosons}
The conserved energy-momentum tensor for gauge vector bosons is
\begin{eqnarray}
T^{\rm V}_{\mu\nu} &=& \eta_{\mu\nu}
\biggl({1\over4} F^{\rho\sigma}
F_{\rho\sigma}-{m^2_A\over2} A^\rho A_\rho\biggr)
- \biggl(F_\mu^{~\rho} F_{\nu\rho} 
- m_A^2 A_\mu A_\nu\biggr)\nonumber\\
&&
-{1\over\xi}\eta_{\mu\nu}\biggl(\partial^\rho\partial^\sigma A_\sigma A_\rho
+{1\over2}(\partial^\rho A_\rho)^2\biggr)
+{1\over\xi}(\partial_\mu\partial^\rho A_\rho A_\nu+
\partial_\nu\partial^\rho A_\rho A_\mu)\ ,
\end{eqnarray}
where the $\xi$-dependent terms correspond to adding a gauge-fixing term $-(\partial^\mu A_\mu -\Gamma^{\mu\nu}_{~~\nu} A_\mu)^2/2\xi$,  with $\Gamma^{\mu\nu}_{~~\nu}=\eta^{\nu\rho}\Gamma^\mu_{~\nu\rho}$ 
the Christoffel symbol.
\begin{center}
\includegraphics[scale=1]{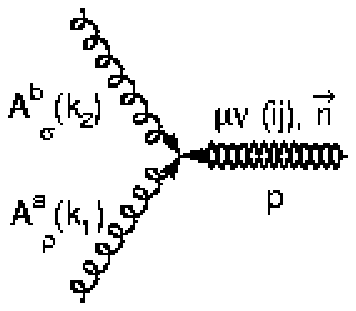}
\end{center}

\begin{eqnarray}
{\tilde{h}}_{\mu \nu}^{\vec{n}}AA&:&\, \, \,-i\, \kappa/2 \, \delta^{ab}((m^2_A + k_1 \cdot k_2)C_{\mu\nu,\rho\sigma}+D_{\mu\nu,\rho\sigma}(k_1 \cdot k_2)+\xi^{-1}E_{\mu\nu,\rho\sigma}(k_1 \cdot k_2))\\
{\tilde{\phi}}^{\vec{n}}_{ij}AA&:&\, \, \,i\,\omega \, \kappa \,\delta_{ij} \delta^{ab}(\eta_{\rho\sigma} m^2_A+\xi^{-1}(k_{1\rho} p_\sigma+ k_{2\sigma} p_\rho))
\end{eqnarray} 

\subsection{Fermions}
The conserved energy-momentum tensor for fermions is
\begin{eqnarray}
T_{\mu\nu}^{\rm F} &=& -\eta_{\mu\nu}
(\overline{\psi}i\gamma^\rho D_\rho\psi-m_\psi\overline{\psi}\psi)
+{1\over2}\overline{\psi}i\gamma_\mu D_\nu\psi
+{1\over2}\overline{\psi}i\gamma_\nu D_\mu\psi\nonumber\\
&&+{\eta_{\mu\nu}\over2}\partial^\rho(\overline{\psi}i\gamma_\rho\psi)
-{1\over4}\partial_\mu(\overline{\psi}i\gamma_\nu\psi)
-{1\over4}\partial_\nu(\overline{\psi}i\gamma_\mu\psi)\ ,
\end{eqnarray}
where we have used the linearized vierbein 
\begin{equation}
e_\mu^{~a}\ =\ \delta_\mu^{~a}
+{\kappa\over2}(h_\mu^{~a}+\delta_\mu^{~a}\phi)\ .
\end{equation}
\begin{center}
\includegraphics[scale=1]{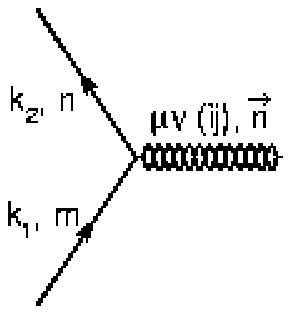}
\end{center}
\begin{eqnarray}
{\tilde{h}}_{\mu \nu}^{\vec{n}}\Psi\Psi&:&\, \, \,-i\, \kappa/8 \, \delta_{mn}(\gamma_\mu(k_{1\nu}+k_{2\nu})+\gamma_\nu(k_{1\mu}+k_{2\mu})-2\eta_{\mu\nu}(k_1+k_2-2m_\Psi)\\
{\tilde{\phi}}^{\vec{n}}_{ij}\Psi\Psi&:&\, \, \,i\,\omega \, \kappa \,\delta_{ij} \delta_{mn}(3/4 k_1+3/4 k_2-2m_\Psi))
\end{eqnarray} 

\section{Confronting with Experiments/Observations}

In the following we will briefly list some of the most interesting constraints on these models. The four principal means of investigating these theories are, 
\begin{enumerate}
\item Deviation from Newton's Law at sub-mm distance 
\item Virtual Graviton Exchange Colliders 
\item Real Graviton production 
\begin{itemize}
\item Missing Energy in Collider Experiments 
\item Missing Energy in Astrophysical sources e.g. Supernovae, Sun, Red Giants 
\item Cosmological consequences e.g. Dark Energy, Dark Matter, Inflation, CMBR 
\end{itemize} 
\item Black hole production at colliders
\end{enumerate} 
We describe only the above mentioned topics. The interested readers can find other useful references here, for the running of couplings and unification in extra dimensions see~\cite{DDG}, for consequences in electroweak precision physics see~\cite{ewprecision}, for neutrino physics with large extra dimensions see~\cite{ADDneutrino}, for topics related to inflation with flat extra dimensions see~\cite{ADDinflation}. Issues related to radius stabilization for large extra dimensions is discussed in~\cite{ADDradion}. Connections to string theory model building can be found in for  example in~\cite{shiutye,pomarolquiros,stringapproach}. 

\subsection{Deviation from Newton's Law at sub-mm distance}

From the relation between the Planck scales of the $(4+n)$ dimensional theory $M_{*}$ and the long-distance 4-dimensional theory $M_{Pl}$,
\begin{equation}
M_{Pl}^2 \sim R^n M_{*}^{n+2}.
\end{equation} 
Putting $M_{*} \sim 1$ TeV then yields
\begin{equation}
R \sim 10^{30/n - 17} \mbox{cm}
\end{equation} 

For $n=1$, $R \sim 10^{13}$ cm, so this case is obviously excluded since it would modify Newtonian gravitation at
solar-system distances. Already for $n=2$, however, $R \sim 1$ mm, which is precisely the distance where our present experimental measurement of gravitational strength forces stops \cite{aharon}. As $n$ increases, $R$ approaches $($TeV$)^{-1}$ distances, albeit slowly: the case $n=6$ gives $R \sim (10$MeV$)^{-1}$. Clearly, while the gravitational force has not been directly measured beneath a millimeter, the success of the SM up to $\sim 100$ GeV implies that the SM fields can not feel these extra large dimensions; that is, they must be stuck on a wall, or ``3-brane", in the higher dimensional space.

\subsection{Virtual graviton exchange}

Besides the direct production of gravitons, another interesting consequence of large extra dimensions is that the exchange of virtual gravitons can lead to enhancement  of certain cross-sections above the SM values.  One can also study the effects of the exchange of virtual gravitons in the intermediate state on experimental observables. Virtual graviton exchange may generate numerous higher dimension operators, contributing to the production of SM particles ~\cite{GRW,HLZ,Hewett,otherADDphen, PRS}.

\subsection{Graviton production in colliders}

Some of the most interesting processes in theories with large extra dimensions involve the production of a single graviton mode at the LHC or NLC. In addition to their traditional role of probing the electroweak scale, they can also look into extra dimensions of space via exotic phenomena such as apparent violations of energy, sharp high-$p_T$ cutoffs and the disappearance and reappearance of particles from extra dimensions. Some of the typical Feynman diagrams for such a process are given in ~\cite{ADDphen,GRW,HLZ,MPP}:

Since the lifetime of an individual graviton mode is of the order $\Gamma \sim \frac{m^3}{M_{Pl}^2}$, which means that each graviton produced is extremely long lived, and once produced will not decay again within the detector. Therefore, 
it is like a stable particle, which is very weakly interacting since the interaction of individual KK modes is suppressed by the four dimensional Planck mass, and thus takes away undetected energy and momentum. This would provide
missing energy signals in accelerator experiments.

\subsection{Supernova cooling}

Some of the strongest constraints on the large extra dimensional scenarios come  from astrophysics. The gravitons are
similar to goldstone bosons, axions and neutrinos in at least one respect. They can carry away bulk energy from an astrophysical body and accelerate its cooling dynamics. These processes have been discussed in detail in \cite{ADDphen,SN}. 

We consider the supernova 1987A. There, the maximum available energy per particle is presumed to be between 20 and 70 MeV .  The production of axions in supernovae is proportional to the axion decay constant $1/f_a^2$. The production of gravitons is also roughly proportional to $1/M_{Pl}^2 (T/\delta m)^n \sim T^n/M_{*}^{n+2}$, where $T$ is a typical temperature within the supernova. This means that the bounds obtained for the axion cooling calculation can be applied using the substitution $1/f_a^2\to  T^n/M_{*}^{n+2}$. For a supernova $T\sim 30$ MeV, and the usual axion bound $f_a \geq 10^9$ GeV implies a bound of order $M_*\geq 10 - 100$ TeV for $n=2$. For $n>2$ one does not get a significant bound on $M_*$ from this process. Of course, when the number of dimensions gets  large enough so that  $1/R \gtrsim 100$ MeV, (corresponding to $n \gtrsim 7$), none of the  astrophysical bounds apply, since all the relevant temperatures would be too low to produce even the lowest $KK$ excitation of the graviton.

For the sun T $\sim$ 1 keV and for red giants T $\sim$ 100 keV . Therefore, even for the maximally dangerous case of weak scale i.e., $1 TeV$ and $n=2$ would be totally safe.

\subsection{Cosmological implications}

Finally we come to the early universe. The most solid aspect of early cosmology, namely primordial nucleosynthesis, remains intact in ADD framework. The reason is simple: The energy per particle during nucleosynthesis is at most a few MeV, too small to  significantly excite gravitons.  Furthermore, the horizon size is much larger than a mm so that the expansion of the universe is given by the usual 4-dimensional Friedmann equations. Issues concerning very early cosmology, such as inflation and baryogenesis may change. This, however, is not necessary since there may be just enough
space to accommodate weak-scale inflation and baryogenesis.

The cosmological models with large extra dimensions offer new ways of understanding the universe\cite{ADDphen}. There exist new  scenarios of inflation and Baryogenesis within the braneworld context. These scenarios manifestly use properties of branes. For instance,  inflation on ``our brane'' can be obtained  if another brane falls on top of ``our brane'' in the early period of development of the brane-universe \cite {DvaliTye}. The potential that is created by another brane in ``our world'' can be viewed as the conventional inflationary potential. Baryon asymmetry of a desired magnitude can also be produced during the collision of these two branes \cite {DvaliGabad}. For more recent developments  see Refs. \cite {E1},\cite {E2}, \cite {E3}, \cite {E4}. With the variants of ADD model, where the extra dimensions comprise a Compact Hyperbolic Manifold, it is possible to solve most of the cosmological problems like homogeneity, 
flatness etc, without inflation \cite{SST}.

Some of the strong constraints come from the fact that at large temperatures, emission of gravitons into the bulk would be a very likely process. This would empty our brane from energy density, and  move all the energy into the bulk in the form of gravitons. To find out at which temperature this would cease to be a problem, one has to compare the cooling rates of the brane energy density via the ordinary Hubble expansions and the cooling via the graviton emission. The two cooling rates are given by
\begin{eqnarray}
\frac{d\rho}{dt}_{expansion}&\sim& -3 H\rho \sim -3 \frac{T^2}{M_{Pl}^2}\rho,\\
\frac{d\rho}{dt}_{evaporation}&\sim& \frac{T^n}{M_*^{n+2}}.
\end{eqnarray} 
These two are equal at the so called ``normalcy temperature'' $T_*$, below which the universe would expand as a normal four dimensional universe. By equating the above two rates we get
\begin{equation} 
T_* \sim \left( \frac{M_*^{n+2}}{M_{Pl}}\right)^{\frac{1}{n+1}} =10^{\frac{6n-9}{n+1}}
\ \ {\rm MeV}.
\end{equation}
This suggests that after inflation, the reheat temperature of the universe should be such that, one ends up below the normalcy temperature, otherwise one would overpopulate the bulk with gravitons and overclose the universe. This is in fact a very stringent constraint on these models, since for example for $n=2$, $T_* \sim 10$ MeV, so there is just barely enough space to reheat above the temperature of nucleosynthesis. However, this makes baryogenesis a tremendously difficult problem in these models. 

\subsection{Black hole production at colliders}

One of the most amazing predictions of theories with large extra dimensions would be  that since the scale of quantum gravity is lowered to the TeV scale, one could actually form black holes from particle collisions at the LHC. Black holes
are formed when the mass of an object is within the horizon size corresponding to the mass of the object. 

What would be the characteristic size of the horizon in such models? This usually can be read off from the Schwarzschild solution which in four dimensions is given by (c=1)
\begin{equation}
ds^2 = (1-\frac{GM}{r}) dt^2-\frac{dr^2}{(1-\frac{GM}{r})} +r^2 d^2 \Omega,
\end{equation}
and the horizon is at the distance where the factor multiplying $dt^2$ vanishes:
$r_H^{4D}= GM$. In $4+n$ dimensions in the Schwarzschild solution the prefactor is 
replaced by $1-\frac{GM}{r} \to 1- \frac{M}{M_*^{2+n} r^{1+n}}$, from which the 
horizon size is given by 
\begin{equation}
r_H \sim \left( \frac{M}{M_*}\right)^\frac{1}{1+n} \frac{1}{M_*}.
\end{equation}
The exact solution gives a similar expression except for a numerical prefactor in the above 
equation. Thus we know roughly what the horizon size would be, and a black hole will form if
the impact parameter in the collision is smaller, than this horizon size. Then the particles
that collided will form a black hole with mass $M_{BH}=\sqrt{s}$, and the cross section as 
we have seen is roughly the geometric cross section corresponding to the 
horizon size of a given collision energy 

\begin{equation}
\sigma \sim \pi r_H^2 \sim \frac{1}{M_{Pl}^2} 
(\frac{M_{BH}}{M_*})^{\frac{2}{n+1}}
\end{equation}
The cross section would thus be of order $1/$TeV$^2\sim 400$ pb, and the LHC would produce
about $10^7$ black holes per year! These black holes would not be stable, but decay via 
Hawking radiation. This has the features that every particle would be produced with an equal
probability in a spherical distribution. In the SM there are 60 particles, out of which there
are 6 leptons, and one photon. Thus about 10 percent of the time the black hole would decay into 
leptons, 2 percent of time into photons, and 5 percent into neutrinos, which would be observed as
missing energy. These would be very specific signatures of black hole production at the
LHC. More detailed description of black hole production can be found in \cite{BH,BHcosmic,otherBH}.

\section{Conclusion}

Even though the extra dimensions look very exotic in the beginning, their inception in the modern physics has helped us a great deal in understanding some of the long standing problems in particle physics and cosmology.
Over the last twenty years, the hierarchy problem has been one of the central motivations for constructing extensions of the SM, with either new strong dynamics or supersymmetry stabilizing the weak scale. By contrast, ADD have proposed that the problem simply does not exist if the fundamental short-distance cutoff of the theory, where gravity becomes comparable in strength to the gauge interactions, is near the weak scale. This led immediately to the requirement of new sub-mm dimensions and SM fields localized on a brane in the higher-dimensional space.  On the other hand, it leads to one of the most  exciting possibilities for new accessible physics,  since in this scenario the structure of the quantum gravity can be experimentally probed in the near future. In summary, there are many new interesting issues that emerge in ADD framework. Our old ideas about unification, inflation, naturalness, the hierarchy problem and the need for supersymmetry are abandoned, together with the successful supersymmetric prediction of coupling constant unification \cite{DG}. Instead, we gain a fresh framework which
allows us to look at old problems in new ways. 

\section*{Acknowledgements}
We would like to thank Dr.Christos Kokorelis for many suggestions and useful correspondence. We thank Rizwan ul Haq Ansari for going through the first draft of this article.

\end{document}